# Spectroscopic Gradients in Elliptical Galaxies

J. Jesús González

*Instituto de Astronomía, U.N.A.M.,
Apdo. Postal 70-264, 04510 México D.F., Mexico*

Javier Gorgas

*Departamento de Astrofísica, Facultad de Ciencias Físicas,
Universidad Complutense de Madrid, 28040-Madrid, Spain*

**Abstract.** Preliminary results based on a large compilation of line-strength profiles are presented and discussed. Elliptical galaxies (Es) with strong central $Mg_2$ show steeper line-strength gradients. These galaxies are more massive and have higher velocity dispersions, note nevertheless, that the $Mg_2$ gradient correlates better with central $Mg_2$ than with any other global or structural parameter. This correlation, and the fact that $Mg_2$ measures $Z/Z_\odot$ together with the [Mg/Fe] abundances among Es, are of extreme importance for the use of the central $Mg_2 - \sigma$ relation as a reflection of a mass-metallicity relation. Spectroscopic gradients arise mainly from a radial decrease in mean stellar metallicity. However, the $Mg_2$ and $<Fe>$ gradients do not correlate within measuring errors, indicating that something else, besides metallicity, is also varying within a galaxy. $H\beta$ gradients indicate that the centers of Es tend to look spectroscopically "younger" than their stellar populations further out.

## 1. Introduction

Spectroscopic gradients are a conspicuous indicator and possibly a measure of the amount of dissipation during the star-forming phase of galaxy formation. They also allow us to study the connection of the chemical and star-formation histories with the local and global properties. Color and line-strength gradients in Es have been detected and measured for over 20 years (see Faber 1977, for an early review), but we have just begun to disentangle their information through the gathering of more and better data together with the development of a more complete theoretical framework. In this talk, we will concentrate on reviewing the preliminary information that the authors have started to gather from the extensive, but scattered and not always homogeneous collection of line-strength profiles from our own research programs and from the literature.

Our intention in this talk is to stress the relevance of spectroscopic gradients to our understanding of elliptical galaxies, and to give you a feeling for how much we can be mislead by using central measurements as an indication of the global properties of elliptical galaxies.

## 2. The spectroscopic data

We recently derived $Mg_2$ gradients for about 150 elliptical galaxies from unpublished data obtained by us, and from a number of $Mg_2$ profiles compiled from the literature (Gorgas, *et al.* 1995, Fisher *et al.* 1995, Sansom *et al.* 1994, Carollo & Danziger 1994a, 1994b, Saglia *et al.* 1993, González 1993, Davies *et al.* 1993, Carollo *et al.* 1993, Davidge 1991a, 1991b, 1992, Bender & Surma 1992, Boroson & Thompson 1991, Gorgas *et al.* 1990, Thomsen & Baum 1989 & 1987, Couture & Hardy 1988, Baum *et al.* 1986, Efstathiou & Gorgas 1985, Faber 1977). The individual $Mg_2$ profiles were transformed to the same reference system (Faber *et al.* 1985) and were projected to the same position angle along the major axis of each galaxy. Relative errors, within and among sources, were also estimated to associate a weight to each data point.

For each galaxy, we derived a gradient of the form $\Delta Mg_2/\Delta \log(r/r_e)$ fitted from $\sim 1.5''$ to the effective radius, $r_e$. Based on the repeatability of multiple observations, the errors in the data and in the fits, as well as on the radial coverage of each galaxy, we ranked the $Mg_2$ gradients by overall quality and reliability. The results discussed here derive from the 114 Es with the most reliable $Mg_2$ gradients in our collected sample.

The $Mg_2$ measurements were complemented with $<Fe>$ (mean of the absorption strength of Fe features at $\lambda 5270$ and $\lambda 5335$) and $H\beta$ data. These *atomic* spectral indices are more affected by differences in resolution (and velocity-dispersion corrections) between sources than are the broader *molecular* indices like $Mg_2$. Therefore, data for these atomic indices are only available, in an homogeneous way, for smaller galaxy samples (e.g., González 1993, Carollo *et al.* 1993, Gorgas *et al.* 1990, Fisher *et al.* 1995, Gorgas *et al.* 1995, Davies *et al.* 1992). $H\beta$ profiles are considered less reliable in the literature given the serious difficulties of nuclear and extended emission so commonly present in elliptical galaxies.

## 3. Metallicity gradients and global parameters

In her pioneering work Faber (1977) discussed the line-strength gradients of early-type galaxies. The gradients are in the sense of decreasing line-strength outwards in a galaxy. For her small sample of bright galaxies the gradient profiles looked quite similar, apparently reflecting an universal decrease of metallicity with properly-scaled radius (e.g., $r/r_e$). During the 80's more and better data were collected, and it was clearly established that spectroscopic gradients show significant variations among Es. Gorgas *et al.* (1990; hereafter GEA) found a hint of a correlation of the metal line-strength gradients with the central $Mg_2$ index and velocity dispersion, but a less clear trend with absolute magnitude.

Even when the $Mg_2$ index (a measure of molecular Mg H and atomic Mg absorption) is not only a metallicity indicator, but also reflects the variations on [Mg/Fe] overabundance *among* Es (e.g., Worthey *et al.* 1992; hereafter WFG), we can still consider this index a good indicator of relative metallicity *within* a galaxy, and equate for the moment the $Mg_2$ gradient to metallicity gradient, delaying further discussion to §5.

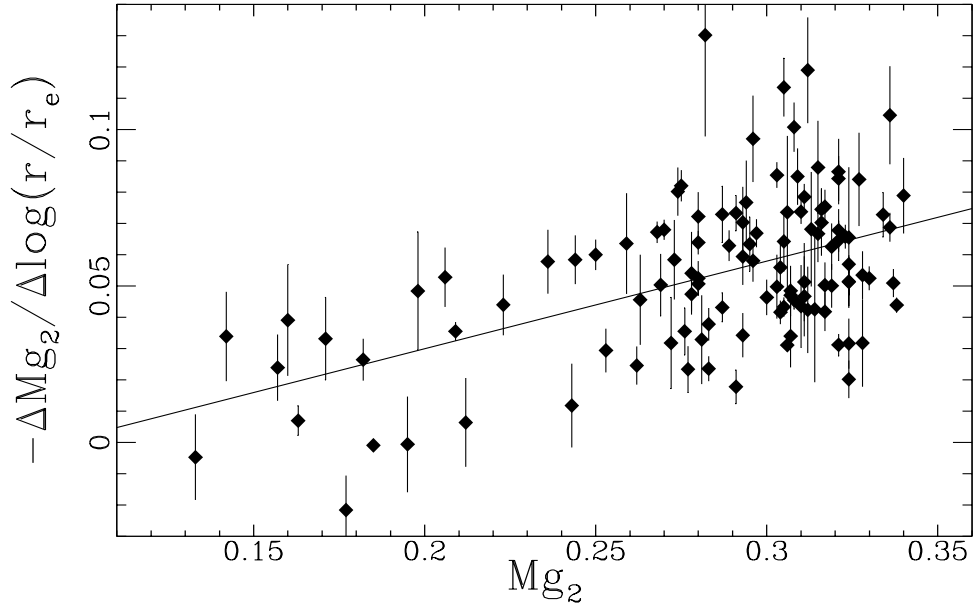

Figure 1. Mg$_2$ Gradient versus central Mg$_2$ for 114 Es with reliable gradients. This relation, $\Delta \mathrm{Mg}_2 / \Delta \log(r/r_e) = 0.026 - 0.28 \mathrm{Mg}_2$, is the best correlation we found of the metallicity gradient with any other global or structural parameter (non-correlation probability, $P_{nc} = 2.8 \times 10^{-5}$, and a RMS scatter of $\sigma = 0.022$ in $\Delta \mathrm{Mg}_2 / \Delta \log(r/r_e)$)

CCD-spectroscopic data of 11 bright Es by Davidge (1992) also show a trend of steeper Mg$_2$ gradients at higher velocity dispersions, luminosities and central line-strengths. His data also suggested that more isotropic galaxies (higher $(v/\sigma)^*$) had shallower gradients than their anisotropic counterparts, but this trend has not been confirmed by other studies (e.g., Fisher et al. 1995, Carollo et al. 1993; hereafter CDB). CDB combine their larger data set with those by GEA and Davies et al. (1993) and define the first clear trends of gradients with global parameters. In particular, they find that the Mg$_2$ gradient steepens with galactic mass, velocity dispersion and central Mg$_2$. Since these correlations are cleaner and more evident for galaxies with $M \lesssim 10^{11} M_\odot$, CDB discussed scenarios in which gradients are driven by dissipation in low-mass Es, but not so for massive galaxies, where other processes like mergers and violent relaxation can play an important role.

Most previous studies are based on limited samples of galaxies. With our large collected sample, we can revise how well the metallicity gradient relates to other global or structural parameters. Given their relative faintness, there is an understandable paucity of spectroscopic data for fainter Es. The present study has been highly enriched by the inclusion of data for 7 dwarf and compact Es recently observed (Gorgas et al. 1995). For the 114 Es with the most reliable gradients in the compiled sample, we confirm the reported loose trends of the Mg$_2$ gradient with galactic mass (CDB), and a slightly better (but far from perfect) correlation with central velocity dispersion than found before.

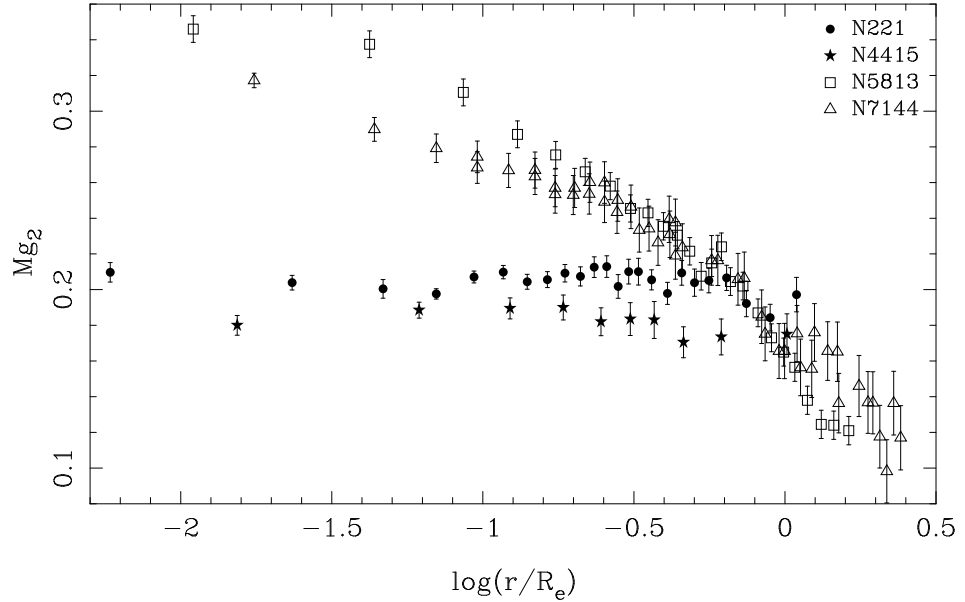

Figure 2. Typical Mg$_2$ profiles for bright (NGC 5813 and NGC 7144) and dwarf (M32 and NGC 4415) elliptical galaxies (data are respectively, from Efstathiou & Gorgas 1985, Saglia *et al.* 1993, González 1993 and Gorgas *et al.* 1995). The global metallicity of bright and low luminosity galaxies may vary much less than inferred from the central line-strengths (see text).

The best correlation we could find for the gradient with any global parameter was with central Mg$_2$ line-strength (Figure 1). From the Spearman coefficient, the probability for no-correlation of these two quantities, $P_{nc} = 2.8 \times 10^{-5}$, is quite significant, in contrast to $P_{nc} = 0.15$ for the correlation with central $\sigma$, $P_{nc} = 0.10$ with (B-V), $P_{nc} = 0.08$ with $\log(r_e)$, $P_{nc} = 0.27$ with absolute blue magnitude, and only $P_{nc} = 0.32$ with galactic mass.

The amount of dissipation during the bulk of the star-formation process may be proportional to the depth of the potential well of a galaxy, setting in a positive correlation of the metallicity gradient with mass, velocity dispersion and central metallicity. Further star-formation, either from cooling flows or triggered by mergers or interactions, would likely be more important in the central parts of a galaxy, blurring the gradient correlation with mass and velocity dispersion. By contrast, the correlation with central Mg$_2$ will be stressed, since gradient and central line-strength are affected in the same sense by more recent star formation at the center. In fact, the central Mg$_2$ index is depressed in Es with signs of a past interaction (Schweizer *et al.* 1990) and in cooling flow galaxies (Cardiel *et al.* 1995) reflecting the presence of a warmer population. A galaxy with even a moderate burst of star formation in its center will also show a shallower-than-before Mg$_2$ gradient due to the decrease in central Mg$_2$. After the warm stars that formed in the burst have evolved, the central line-strength and the gradient will increase again, at least to their original pre-burst values, or even to higher values due to the presence now of a new oldish metal-enriched population of stars. For the moment, we can qualitatively understand in these simple terms that the $Mg_2$ gradient correlates better with central $Mg_2$ than with any other parameter.

To illustrate the limitation of central $Mg_2$ as a measure of global metallicity, Figure 2 shows a pair of the relatively steep $Mg_2$ profiles characteristic of bright elliptical galaxies, and a couple of flatter profiles typical of dwarf Es. Flat and steep profiles crossover within or around an effective radius, while the line-strength of bright Es keeps on decreasing at large radii, reaching $Mg_2$ values comparable to galactic globular clusters (e.g., Burstein *et al.* 1984). So far, no steep $Mg_2$ profile has shown signs of flattening out even beyond 2 $r_e$.

Galaxies with higher central $Mg_2$ are also brighter, less dense, more massive, and with higher and more anisotropic velocity dispersions. Furthermore, they have steeper gradients, as discussed above. We have to keep this in mind when interpreting relations involving line-strengths with global parameters, and be aware of the limitations of central line-strength as an indicator of global metallicity.

## 4. Gradients and global relations: the $Mg_2 - \sigma$ relation, an example

The well known positive correlation between central $Mg_2$ and velocity dispersion (e.g., Burstein *et al.* 1988) has been interpreted as a reflection of a mass-metallicity relation. Such a relation can be naturally expected from scenarios of purely dissipative collapse or from less dissipative scenarios coupled with supernovae induced winds (e.g., Larson, 1974, Carlberg 1984, Arimoto & Yoshii 1987, Matteucci and Tornabè 1987, Brocato *et al.* 1990, Stiavelli & Matteucci 1991).

The central $Mg_2 - \sigma$ relation may not be the best indicator of a cosmic mass-metallicity relation. In fact, $Mg_2$ varies among Es more than expected from other metallicity indicators like $< Fe >$, suggesting that the [Mg/Fe] abundance ratio is higher than solar in brighter Es, probably due to an enrichment dominated by type-II supernovae (WFG, González 1993, Davies in this volume). $Mg_2$ is then a measure of both, $Z/Z_\odot$ and [Mg/Fe]. Among Es, the stellar metallicity not only may be varying less than expected by the observed range in central $Mg_2$, but the $Mg_2$ gradient also correlates with velocity dispersion and central line-strength (§3). Therefore, the cosmic mass-metallicity relation should be considerably shallower than inferred from taking central $Mg_2$ as a pure metallicity indicator or as a good relative measure of global abundance (Figure 3).

### 4.1. Residuals from the $Mg_2 - \sigma$ relation

The scatter in the $Mg_2 - \sigma$ relation has been discussed at length by Burstein *et al.* (1988) and more recently by Bender *et al.* (1992). $Mg_2$ deviations at a fixed $\sigma$ can be due to mergers and interactions and/or a combination of age and metallicity effects, but also to some extent, to the partial information that the central line-strength reflects about the overall metallicity of a galaxy.

Figure 4 shows that the $Mg_2 - Mg_2(\sigma)$ residual in fact correlates with the $Mg_2$ gradient ($P_{nc} = 2.6 \times 10^{-5}$), but this is far from being the end of the story: although some outliers are removed, the intrinsic scatter of the gradient-corrected $Mg_2 - \sigma$ relation (not shown) is not significantly lower than the in-trinsic scatter from the original relation. We would expect the gradient to help reduce residuals, at least in principle, because the total observed scatter from the $Mg_2(r_e) - \sigma$ relation is not significantly larger than for the central $Mg_2$ values, (.032 and .030 mag in $Mg_2$ respectively), as shown in Figure 3, despite the larger measuring errors and lower line-strengths at one effective radius. Nevertheless, the non-significant reduction of scatter in a gradient-corrected $Mg_2 - \sigma$ relation is not surprising after all, because an anomalous nucleus will also affect the observed line-strength profile, causing the gradient, as measured, to over correct for aperture effects.

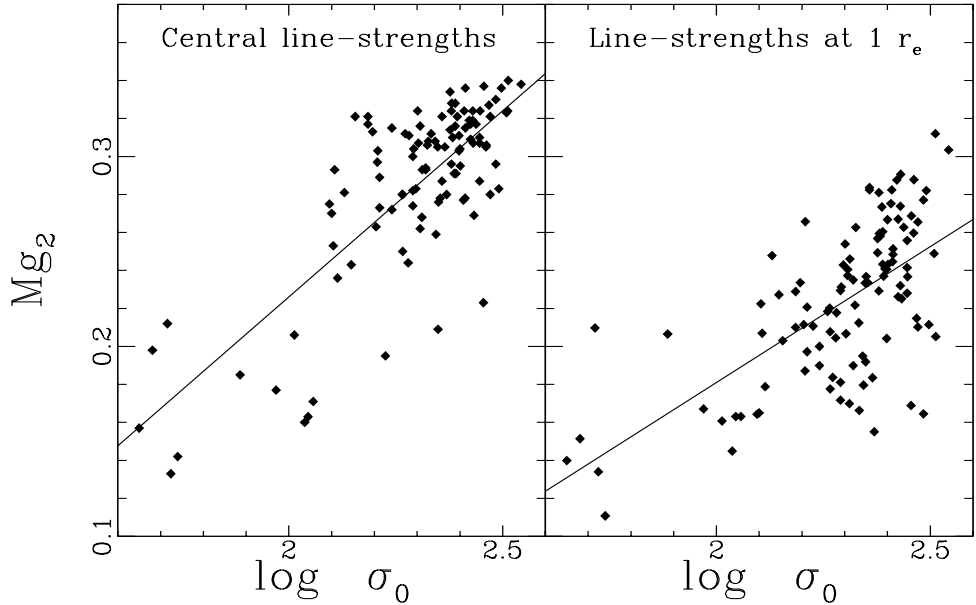

Figure 3. $Mg_2 - \sigma_0$ relation for central $Mg_2$ (left) and $Mg_2(r = r_e)$ (right) for the Es in Figure 1. The relation on the left, $Mg_2 = -0.166 + 0.196 \log(\sigma_0)$, is consistent with those derived by Burstein et al. (1988) and Bender et al. (1993). The relation is flatter for the line-strengths at the effective radius ($Mg_2(r_e) = -0.105 + 0.143 \log(\sigma_0)$). Bright and faint Es have more similar metallicities at $\sim r_e$ than at their nuclei. The intrinsic relation of mass and *global* metallicity should be flatter than expected from using *central* $Mg_2$ as global-metallicity indicator. Interestingly enough, after accounting for measuring errors, the intrinsic scatter of the $Mg_2(r_e) - \sigma_0$ relation is not higher (or even smaller) than from the central $Mg_2 - \sigma_0$ relation.

An independent clue for understanding the scatter from the $Mg-\sigma$ relation comes from the $H\beta$ index. Figure 4 also shows the anti-correlation of the $Mgb - Mgb(\sigma)$ residuals with the central $H\beta$ absorption for González's (1993; hereafter G93) sample of "normal" elliptical galaxies. The atomic $Mgb$ index is basically equivalent to the molecular $Mg_2$ index, $Mg_2 \simeq 0.066$ EW($Mgb$), but is more accurately measured in G93. Most of the galaxies in this sample are field galaxies or members of small groups, showing no clear signs of distortion or interaction, and some were corrected for the effect of moderately low levels of emission.

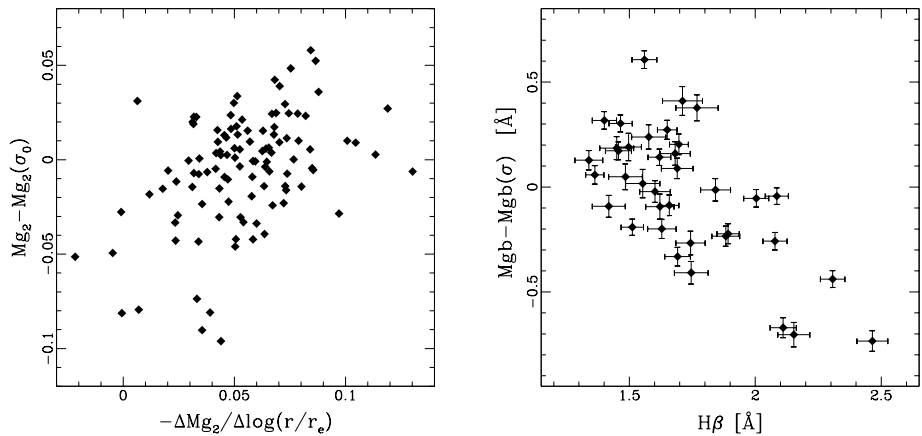

Figure 4. $Mg_2 - Mg_2(\sigma_0)$ residuals as a function of the $Mg_2$ gradient, for the sample of 114 Es with reliable $Mg_2$ profiles (left). Although the residuals correlate with the gradient, applying a gradient correction to the $Mg_2-\sigma$ relation does not reduce its scatter significantly. The figure on the right shows the correlation of the central H$\beta$ absorption with Mgb residuals from the Mgb-$\sigma$ relation, now for a sample of 37 *normal* Es (González 1993). The cosmic scatter from the Mgb-$\sigma$ relation of relatively-undistorted bright Es ($\sim 6\%$) gets substantially reduced ($\lesssim 4\%$) after correction for central H$\beta$ absorption.

Accounting for observational errors, the intrinsic scatter of the Mgb $-\sigma$ relation is 0.28 Å in the equivalent width of Mgb (or about 6%), for the 35 brightest galaxies in G93. After correcting the relation using the correlation of its residuals with central H$\beta$ absorption, the intrinsic scatter is significantly reduced (to 0.18 Å in EW(Mgb), or $\sim 3.8\%$), even for this sample of relatively unperturbed Es. Probably, the mechanism that drives the strong range in H$\beta$ absorption among Es, is the same that drives substantial scatter in the Mg-$\sigma$ relation (spectroscopic mean-age of the stellar populations?).

## 5. Pure metallicity gradients?

Despite the clear differences in [Mg/Fe] among Es, in this talk we have considered $Mg_2$ as a good relative indicator of metallicity within a galaxy. This is so because the Mg and Fe line-strengths change in proportion to each other within a galaxy (GEA, G93, WFG, CDB), consistent with a pure metallicity change as expected from the relation defined by globular clusters (Burstein *et al.* 1984) and by the nearly-degenerated age-metallicity relation defined by population synthesis models (Worthey 1994; hereafter W94).

Nevertheless, Mg and Fe gradients do not correlate within the errors, as shown by G93 from Mgb and $<Fe>$ data (Figure 5). This conclusion is also supported by CDB and GEA's data. Spectroscopic gradients must then define at least a two-parameter family, where metallicity can be identified as their main driver. Abundance ratios may also be varying within Es but very high signal-to-noise data would be required to study second order effects in Mg-Fe diagrams. Another possibility are variations in mean stellar age of the populations inside a galaxy.

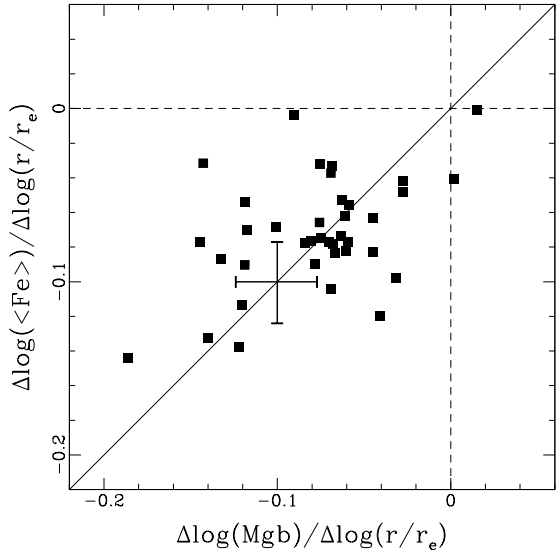

Figure 5. Mgb and $<Fe>$ gradients for the G93 sample of normal Es (typical error bars are shown). Both line-strength indices are expected to decrease nearly in proportion to one another (close to the unity-slope line in the diagram) if only the mean metallicity of the stellar population diminished radially inside Es. The scatter in this diagram is larger than expected from observational errors, indicating that some other stellar-population parameter, besides metallicity, must also be varying radially inside an elliptical galaxy (see text).

Gradients in mean-stellar age within Es, in the sense of "younger" populations towards the center, are supported by the H$\beta$ profiles. G93 finds a flat mean-gradient in H$\beta$ for his sample of normal-looking Es. This observation is also supported by GEA's data and by observations in bright cluster members (Fisher *et al.* 1995). The H$\beta$ absorption of an old stellar population is expected to increase with decreasing metallicity as the red-giant branch, and the main sequence turn-off, move to higher temperatures. Therefore, we need to call on a mechanism that reduces the H$\beta$ absorption outwards, to compensate for the effect of decreasing metallicity within a galaxy. According to G93 data and W94 age-metallicity calibration, a moderately small gradient in mean-spectroscopic age ($\sim 3$Gyr per dex in radius) suffices to keep H$\beta(r) \simeq const$ within a typical elliptical galaxy. Other effects on the spectroscopic gradients, besides internal variations in mean metallicity and mean-stellar age of the populations inside elliptical galaxies, need to be explored.

## 6. Conclusions

Elliptical galaxies show a considerable range in $\log(Z/Z_\odot)$ gradients, ranging from zero or as small as $\sim -.15$ dex per decade in radius, to as much as $\sim 4-5$ times steeper than that (based on the $Mg_2$ calibration of W94). Smaller, fainter, low-central $Mg_2$ galaxies show shallower gradients than brighter Es. The gradients imply that the intrinsic mass-metallicity relation among Es is flatter than that inferred from a simple interpretation of the observed central $Mg_2 - \sigma$ relation. Actually, dwarfs and giant Es may have essentially the same total or *global* metallicity, but metals are less concentrated in smaller galaxies.

The scatter from the central $Mg_2 - \sigma$ relation may, at least in part, be a result from recent episodes of star formation more prominent toward the center of the galaxies. This is inferred from the correlation of the $Mg_2$ residuals with the $Mg_2$ gradient and, more importantly, with $H\beta$ absorption, but more detailed work needs still to be done.

The spectroscopic evidence of elliptical galaxies, both among and within galaxies, points toward the conclusion that bright and faint Es had different chemical enrichment and dissipation during their star-formation histories.

At least qualitatively, we could understand the observed spectroscopic trends in elliptical galaxies if there was a physical process, presently unknown, that strongly coupled dissipation and star-formation efficiency (probably through velocity dispersion) *during the time the bulk of the stars was formed*. Galaxies with short time scales for star formation i.e., forming the bulk of their stars faster and at an earlier epoch, would show a positive [Mg/Fe] abundance ratio. In the scenario where these galaxies also dissipated the most, they would end up, in addition, with a strong metallicity gradient. On the other hand, galaxies with longer star-formation time scales and less dissipation, will show shallower gradients, lower central-metallicities, about solar [Mg/Fe], and younger-looking populations. In both cases, the global metallicity would be comparable (of the order of the yield if most of the original gas is consumed), but with the metals more concentrated towards the center in the systems where the gas dissipated the most. In this scenario, the mass-density anti-correlation of Es would have to be primordial i.e., relative to the era of significant star formation.

**Acknowledgments.** This project is partially supported by DGAPA-UNAM (grant No. IN107094), and by the UNAM-UCM Academic Exchange Program. J. Gorgas acknowledges financial support from the Spanish "Programa Sectorial de Promoción General de Conocimiento" under grant No. PB93-456.